\documentclass[lettersize,journal]{IEEEtran}
\usepackage{amsmath,amsfonts}
\usepackage{algorithmic}
\usepackage{algorithm}
\usepackage{array}
\usepackage[caption=false,font=normalsize,labelfont=sf,textfont=sf]{subfig}
\usepackage{textcomp}
\usepackage{rotating}
\usepackage{stfloats}
\usepackage{url}
\usepackage{verbatim}
\usepackage{graphicx}
\usepackage{cite}
\begin{document}

\title{Q-RESTORE: Quantum-Driven Framework for Resilient and Equitable Transportation Network Restoration}



\author{Daniel Udekwe, Ruimin Ke,~\IEEEmembership{Member, IEEE,}, Jiaqing Lu, Qian-wen Guo

\thanks{Daniel Udekwe, Jiaqing Lu and Qian-wen Guo are with the Department of Civil and Environmental Engineering, FAMU-FSU College of Engineering, Tallahassee, Florida, 32310, USA (email: dau24@fsu.edu; jl23br@fsu.edu; qguo@fsu.edu)}
\thanks{Ruimin Ke is with the Department of Civil and Environmental Engineering, Rensselaer Polytechnic Institute, Troy, New York, 12180, USA. (email: ker@rpi.edu)}
\thanks{Manuscript received xxx, 2025; revised xxx, 2025.}}

\markboth{}%
{Shell \MakeLowercase{\textit{et al.}}: A Sample Article Using IEEEtran.cls for IEEE Journals}


\maketitle

\begin{abstract}
Efficient and socially equitable restoration of transportation networks post-disasters is crucial for community resilience and access to essential services. The ability to rapidly recover critical infrastructure can significantly mitigate the impacts of disasters, particularly in underserved communities where prolonged isolation exacerbates vulnerabilities. Traditional restoration methods prioritize functionality over computational efficiency and equity, leaving low-income communities at a disadvantage during recovery. To address this gap, this research introduces a novel framework that combines quantum computing technology with an equity-focused approach to network restoration. Optimization of road link recovery within budget constraints is achieved by leveraging D-Wave's hybrid quantum solver, which targets the connectivity needs of low-, average-, and high-income communities.
 This framework combines computational speed with equity, ensuring priority support for underserved populations. Findings demonstrate that this hybrid quantum solver achieves near-instantaneous computation times of approximately 8.7 seconds across various budget scenarios, significantly outperforming the widely used genetic algorithm (GA). It offers targeted restoration by first aiding low-income communities and expanding aid as budgets increase, aligning with equity goals. This work showcases quantum computing's potential in disaster recovery planning, providing a rapid and equitable solution that elevates urban resilience and social sustainability by aiding vulnerable populations in disasters.

\end{abstract}


\section{Introduction}
\setlength\parindent{0pt} Urban transportation networks are essential for the economic vitality, social cohesion, and environmental sustainability of modern cities \cite{mouratidis2021transportation, canales2017connected, banister2008sustainable}. These networks facilitate the movement of people and goods, reduce traffic congestion, and lower greenhouse gas emissions, playing a critical role in the functioning of urban environments \cite{adorno2018ageing}. As urban populations grow, with estimates suggesting that 68\% of the global population will reside in cities by 2050, the demand for reliable transportation infrastructure is set to increase substantially \cite{duranton2020economics}. Managing transportation networks under these conditions involves complex challenges such as real-time route optimization, congestion control, dynamic scheduling, and adaptive resource allocation \cite{zahid2023future, mouratidis2022bike, xu2022urban, porru2020smart, cruz2020mobility}. Traditional computational methods, though effective for limited problems, often struggle with the data complexity and interdependencies inherent in optimizing urban transit and shared mobility systems, especially on a large scale \cite{rodrigue2020geography, kumari2010survey}. Moreover, the effects of climate change, urbanization, and mobility demands make these transportation systems increasingly susceptible to disruptions, particularly during natural disasters, where resilience and swift recovery become crucial  \cite{servou2023data}.

\setlength{\parindent}{.15in} In recent years, the frequency and severity of natural disasters have intensified, placing substantial pressure on urban infrastructure. Since 1980, the United States has endured over 400 significant weather and climate disasters, each resulting in damages of \$1 billion or more (adjusted to 2024 dollars) \cite{noaa2023disasters}. The cumulative costs of these events now exceed \$2.785 trillion, underscoring the escalating financial impact and the critical need for resilient urban infrastructure \cite{noaa2023disasters}. For hurricane-prone cities, these events severely impact transportation networks, making post-disaster restoration a critical task for resuming mobility, reconnecting communities, and supporting emergency response \cite{banerjee2023data, neelam2020scientometric}. Traditional restoration methods are typically reactive and constrained by computational limitations, making them inefficient for large-scale disruptions \cite{fang2019optimum}. The extended recovery timelines and inequitable distribution of restoration efforts often leave low-income neighborhoods underserved and isolated, with profound social and economic consequences \cite{tajik2021heuristic}. Given these challenges, there is an urgent need for restoration strategies that are both efficient and equitable, ensuring that vulnerable populations are prioritized in the recovery process. 

Quantum computing, a novel paradigm rooted in the principles of quantum mechanics, offers transformative solutions for complex optimization problems that are infeasible for classical computers \cite{hossain2023potential, nielsen2010quantum, cerezo2021variational, sood2023quantum}. Quantum systems exploit superposition, entanglement, and quantum annealing to handle vast datasets and complex interdependencies, providing advantages for applications in cryptography, machine learning, and industrial information integration \cite{preskill2018quantum, ogur2023effect, lu2023quantum, cho2021quantum, dixit2023quantum, zhuang2024quantum}. In particular, quantum computing shows potential for solving the computationally intensive tasks involved in urban infrastructure management, which could significantly enhance resilience planning and disaster recovery \cite{bonab2023urban}. This study explores the application of quantum computing to the equitable restoration of transportation networks by leveraging D-Wave’s hybrid quantum solver, compared with a traditional optimization method such as genetic algorithm. Genetic algorithms, inspired by evolutionary biology, are commonly used to solve complex optimization problems by iteratively improving solutions through selection, crossover, and mutation \cite{sivanandam2008genetic, kim2016multiobjective}. However, as problem complexity grows, genetic algorithms encounter challenges in scalability and efficiency, limiting their applicability in high-stakes, time-sensitive disaster recovery scenarios \cite{moser2000scalability}. 

A key contribution of this work is the integration of equity as a guiding principle within the quantum-based restoration framework. Equity in post-disaster restoration is often overlooked, with resources typically allocated based on economic or infrastructural considerations rather than the needs of underserved communities \cite{meerow2019social}. By prioritizing the recovery of transportation links that serve low-income areas, our approach aims to ensure that vulnerable populations receive prompt and effective support, reducing the social disparities that often accompany disaster recovery \cite{litman2017evaluating, stacy2020access}. 

This paper introduces a novel framework, Q-RESTORE (Quantum-based Resilient and Equitable Restoration), which combines quantum computing capabilities with an equity-focused optimization strategy to address the restoration of transportation networks. The Q-RESTORE framework uses D-Wave’s hybrid solver to optimize restoration planning, balancing computational efficiency and social fairness in resource allocation. The empirical analysis assesses the computational efficiency and solution quality of Q-Restore by comparing it with a genetic algorithm-based approach across various budget levels. Using a genetic algorithm (GA) exclusively for comparison is justified based on its suitability for solving complex, nonlinear, and multimodal optimization problems, where classical solvers like CPLEX or Gurobi may be computationally infeasible or overly restrictive in problem formulation \cite{saghand2022exact}. GAs offer robust exploration of the search space through stochastic processes, reducing the risk of premature convergence common in alternatives like simulated annealing (SA) or particle swarm optimization (PSO) \cite{katoch2021review}. Their flexibility to integrate domain-specific knowledge through custom fitness functions and genetic operators makes them particularly effective for Constrained Quadratic Models (CQMs) \cite{sampson1976adaptation, mitchell1998introduction, golberg1989genetic}. 

Results indicate that the hybrid quantum solver achieves substantial improvements in computational time, consistently completing restoration tasks within a few seconds, compared the substantial amount of time required by the genetic algorithm for each generation of solution. Additionally, Q-Restore prioritizes restoration efforts for low-income areas, ensuring an equitable allocation of resources that addresses the needs of vulnerable communities. These findings underscore the potential of quantum computing not only to enhance the speed and scalability of disaster recovery but also to promote fairness and resilience in urban infrastructure planning. By advancing the capabilities of post-disaster restoration, Q-Restore offers a powerful tool for urban planners and policymakers to build resilient, inclusive cities in the face of increasing environmental challenges. 

The remainder of this paper is structured as follows: Section \ref{lit.rev.} offers an overview of quantum computing, including its basic principles, common quantum algorithms, and the current state of the field as well as a review on quantum computing in equitable restoration plan optimization and network design planning. Section \ref{methodology} covers the problem formulation, a description of the data source and proposed methods. The results and discussion from both traditional (GA) and D-Wave's hybrid quantum approaches are presented in Section \ref{results}, followed by the conclusion in Section \ref{conclusion}.

\section{LITERATURE REVIEW} \label{lit.rev.}
\setlength\parindent{0pt} This section introduces quantum computing, delving into its fundamental principles, distinguishing it from classical computing, and discussing common algorithms utilized by quantum computers. Additionally, it examines the current state of quantum computing and reviews its application in restoration plan optimization and Network Design Planning (NDP).
\subsection{Basic Principles of Quantum Computing}
\begin{figure}[!t]
  \centering
  \includegraphics[width=0.5\textwidth]{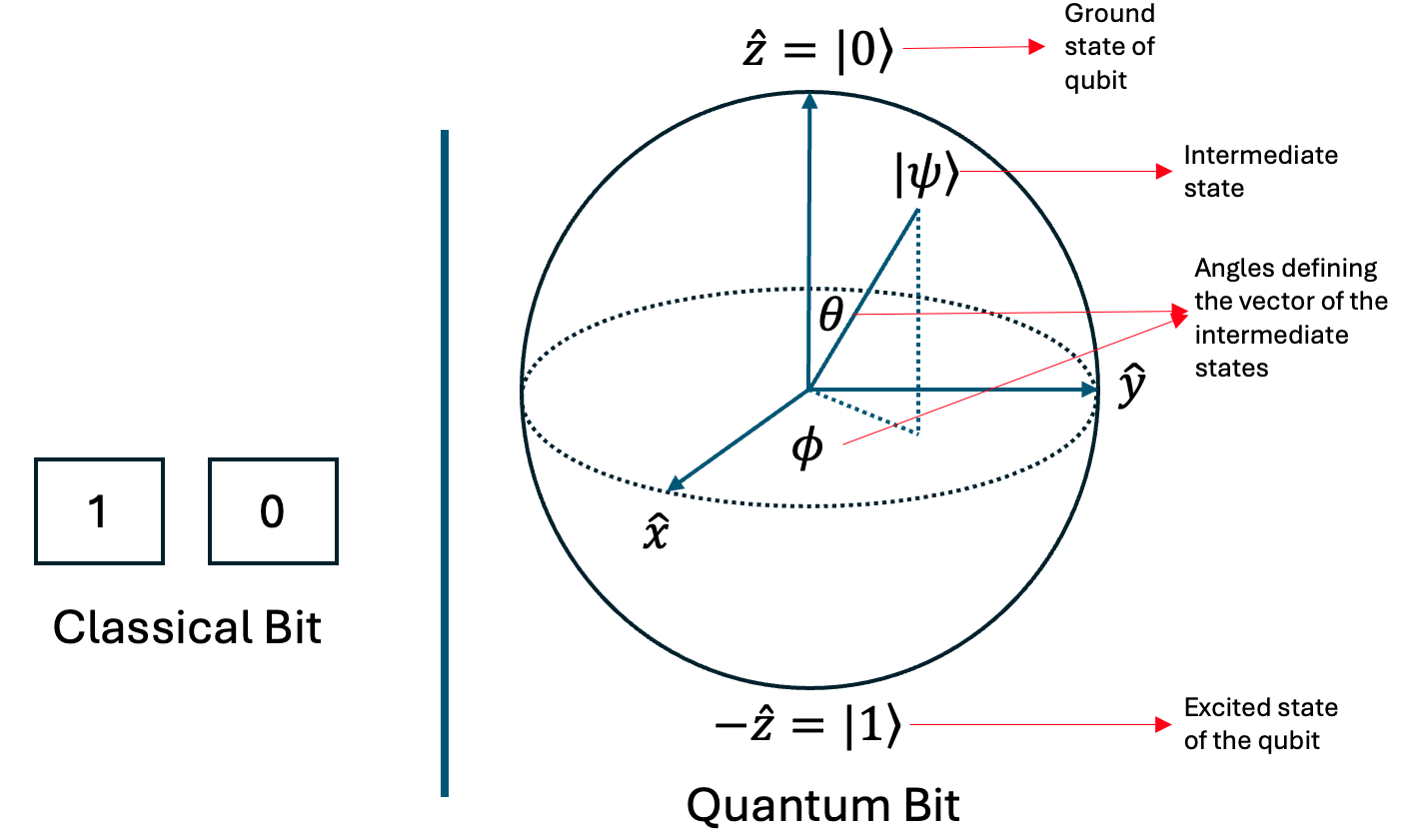}
  \caption{Fundamental units in classical and quantum computing are the bit and qubit}\label{fig:qubit}
\end{figure}

Quantum computing leverages quantum mechanics to process information in ways classical computers cannot \cite{horowitz2019quantum, mccaskey2018language}. Central to this paradigm are qubits, which, unlike classical bits that are either 0 or 1, can exist in a superposition of both states simultaneously \cite{dragoman2004analogies, schroter2022quantum}. This property, illustrated by the \textit{Bloch sphere} in Figure \ref{fig:qubit}, allows quantum computers to process vast amounts of information concurrently. While classical computers handle bits sequentially, quantum computers use superposition and entanglement to perform multiple calculations at once, enhancing computational power and efficiency for complex problems.

\setlength{\parindent}{.15in} Mathematically, a qubit's state can be represented as:

\begin{equation}
    \underbrace{|\psi\rangle}_{\substack{\text{State of}\\ \text{the qubit}}} = \underbrace{\alpha|0\rangle}_{\substack{\text{Probability}\\ \text{of 0 state}}} + \underbrace{\beta|1\rangle}_{\substack{\text{Probability}\\ \text{of 1 state}}},
\end{equation}
\\
where $\alpha$ and $\beta$ are complex numbers that satisfy the normalization condition \cite{nielsen2010quantum}:

\begin{equation}
    |\alpha|^2 + |\beta|^2 = 1.
\end{equation}

\setlength{\parindent}{.15in} Entanglement is a key principle of quantum computing, where interconnected quantum particles influence each other's states instantly, regardless of distance \cite{weinfurter2005power, horodecki2009quantum}. This property allows entangled qubits to perform coordinated operations, significantly enhancing computational power and enabling complex problem-solving \cite{li2023entanglement}. Quantum gates, which manipulate qubits through unitary transformations, form the building blocks of quantum circuits, akin to classical logic gates. Common quantum gates include the Pauli-X, Hadamard, and CNOT gates \cite{barenco1995elementary}.

Building on these foundational elements, quantum computing algorithms harness the unique properties of quantum mechanics—such as superposition, entanglement, and quantum interference—to solve certain problems more efficiently than classical algorithms. Key algorithms include Shor's Algorithm \cite{shor1994algorithms}, Grover's Algorithm \cite{grover1996fast}, Quantum Fourier Transform \cite{hales2000improved}, Quantum Phase Estimation \cite{dorner2009optimal}, Variational Quantum Eigensolver (VQE) \cite{liu2019variational}, Quantum Approximate Optimization Algorithm (QAOA) \cite{farhi2014quantum}, and Quantum Walk Algorithms \cite{ambainis2007quantum}. Additional notable algorithms include the Harrow-Hassidim-Lloyd algorithm \cite{shao2018reconsider}, Boson Sampling \cite{tillmann2013experimental}, and Quantum Annealing \cite{morita2008mathematical}, each designed to exploit the distinctive advantages of quantum computation for diverse applications \cite{parekh2021quantum}.

\subsection{Current State of Quantum computing}

\setlength\parindent{0pt} Quantum computing has seen rapid advancements in both hardware and software over the past decade \cite{coccia2022evolution, sood2023systematic}. Companies like Google and IBM have achieved critical milestones, such as Google's 2019 claim of quantum supremacy with their Sycamore processor, which performed a specific computation faster than the most powerful classical computers \cite{arute2019quantum}. IBM offers cloud-accessible quantum processors, enabling global research collaboration \cite{santos2016ibm}, while Microsoft focuses on scalable quantum technologies and provides the Quantum Development Kit (QDK) for developers \cite{article}. D-Wave Systems specializes in commercial quantum annealers \cite{koshka2020comparison}, and Rigetti Computing is known for full-stack quantum solutions and hybrid systems \cite{de2022towards, hassija2020present}. Despite progress, challenges like error correction and scalability remain, as quantum computers are susceptible to errors from decoherence and noise, and scaling up processors requires advanced engineering \cite{shor1995scheme, kempe2006approaches, byrd2003combined, devoret2013superconducting, bravyi2022future, de2021materials}. Additionally, developing robust quantum algorithms and software is an ongoing research area \cite{self2021variational, haner2018software, gill2024quantum}. Though in its early stages, the field's rapid advancements suggest a promising future, potentially transforming various industries.

\subsection{Quantum Computing in Network Design Planning}
\setlength\parindent{0pt} The Network Design Problem (NDP) involves optimizing the configuration and management of transportation networks, which presents computational challenges due to the complexity and scale of real-world networks. Traditional methods often fall short in handling the computational demand of large-scale NDP, resulting in slow or even infeasible solutions. Quantum computing offers substantial advantages for addressing these challenges, as its parallelism and superposition capabilities allow for exploration of vast solution spaces far more efficiently than classical computing methods \cite{cooper2021exploring, lin2021intelligent}. While the field is still emerging, researchers have begun exploring quantum algorithms and approaches to achieve breakthroughs in solving complex network design and optimization problems, which could revolutionize the efficiency and scalability of solutions for the NDP \cite{lin2009quantum, dixit2023quantum, lin2019dual}.

\setlength{\parindent}{.15in} The Restoration Plan Optimization (RPO) problem, a specific type of NDP, focuses on optimizing transportation network restoration in disaster contexts, where infrastructure resilience and rapid recovery are critical \cite{orabi2010optimizing}. In this context, road capacity degradation due to infrastructure damage increases the complexity of the problem, as restoration plans must not only address mobility disruptions but also account for additional uncertainties and constraints. Quantum computing shows great promise for enhancing RPO, as quantum-inspired algorithms, such as those proposed by \cite{lin2009quantum}, can explore large solution spaces efficiently to improve restoration effectiveness. For instance, \cite{zhang2020application} demonstrated quantum genetic optimization’s potential in traffic network prediction, which is beneficial for RPO by forecasting traffic patterns and vulnerabilities. Furthermore, dual approximation-based quantum-inspired algorithms, like those from \cite{lin2019dual}, adapt dynamically to changing conditions, ensuring resilient and adaptable restoration plans. This quantum integration ultimately enhances the resilience of transportation networks, offering a robust framework to support quick and efficient recovery in the face of disruptions.
\section{METHODOLOGY} \label{methodology}
\setlength\parindent{0pt} This section discusses the formulation of the equitable restoration optimization problem, presents the data source used for the Sioux Falls area in this study, and discusses both the genetic algorithm and quantum approaches.
\subsection{Problem Formulation} 
\begin{figure}[!t]
  \centering
  \includegraphics[width=0.4\textwidth]{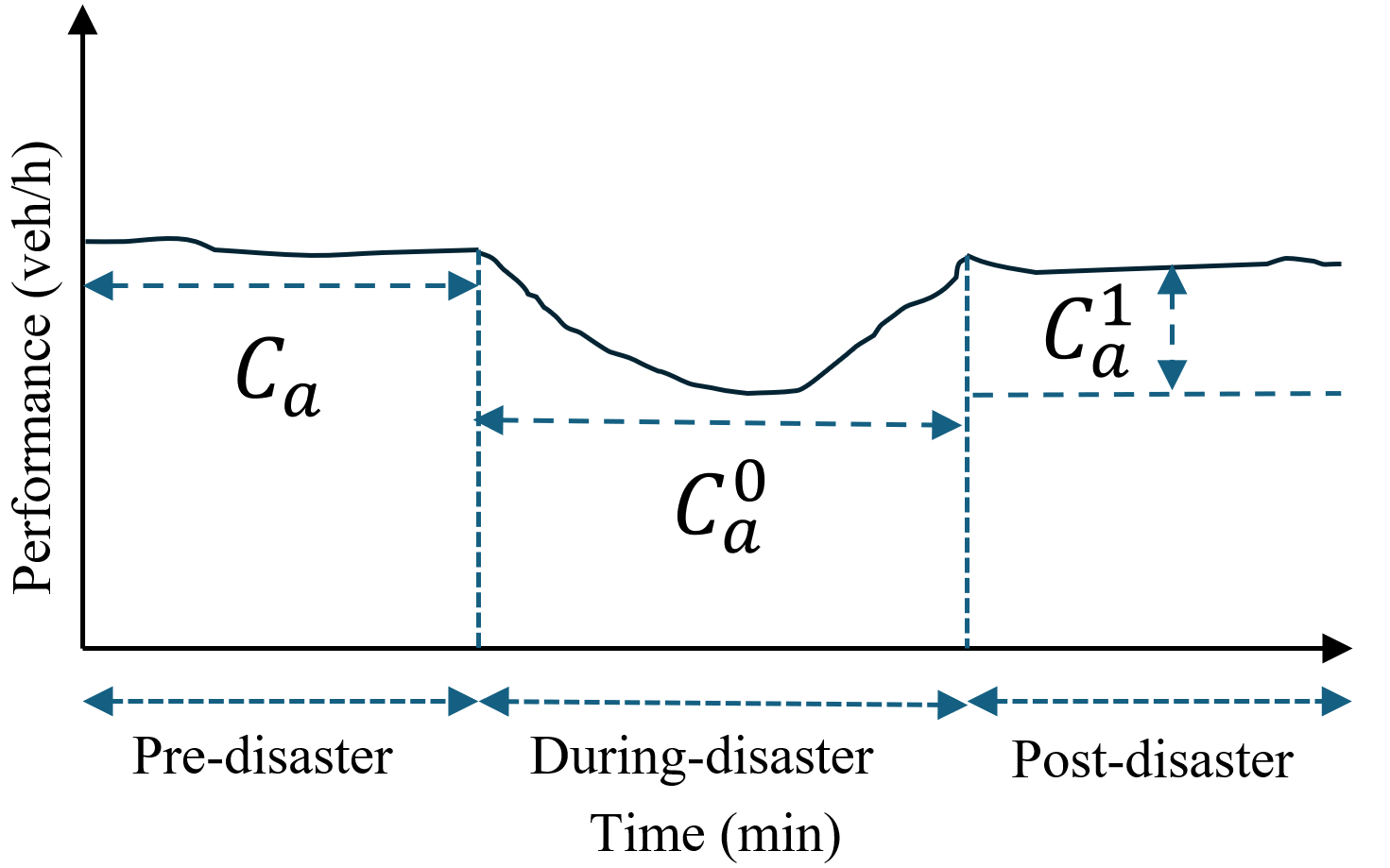}
  \caption{Link capacity variation before, during, and after a disaster event}\label{fig:link_varia}
\end{figure}

\setlength\parindent{0pt} In this study, the restoration plan optimization after a disruptive event is formulated as a bi-objective bi-level optimization. Bi-level optimization is often referred to as a ``leader-follower problem". The ``leader" represents the upper-level decision maker who makes decisions that affect the ``follower", who represents the lower-level decision-maker. The follower's response to the leader's actions determines the overall outcome of the problem. The notation used in this formulation is detailed in Table \ref{tab:notation} while a description of the link capacity variation before, during, and after a disaster event is given in Figure \ref{fig:link_varia}.

\begin{table}[!t]
    \caption{Symbols and definitions used all through this study}\label{tab:notation}
    \begin{center}
        \begin{tabular}{l l c}
            \hline
            Symbol& Definition & Unit \\\hline
            $R$   & Resilience Measure & - \\
            $\mu$ & Weight with range [0,1] & - \\
            $B$    & Budget Size    &   $veh/h$      \\ 
            $C$     & Maximum capacity recovery size & $veh/h$ \\
            $D$   & Mobility Measure  & -  \\
            $E$   & Equity Measure & -  \\
            $M_a(c_a^1)$ & Cost of restoring link $a$ with capacity $c_a^1$ & - \\
            $C_a$  & Initial link capacity before hurricane & $veh/h$  \\
            $C_a^0$ & Capacity of link $a$ after hurricane-induced hazard & $veh/h$ \\
            $C_a^1$ & Capacity improvement for link $a$ & $veh/h$ \\
            $x_a$  & Traffic flow on link $a$ & $veh/h$ \\
            $t_a$ & Travel time on link $a$ & $min$ \\
            $f_{rs}^k$ & Traffic flow on path k connecting  & $veh/h$\\
                        & node $r$ (origin) and node $s$ (destination) &\\
            $\delta_{rs}^{a,k}$ & Binary indicator variable & -  \\
            $k_{rs}$  & Number of candidate paths from node $r$ to $s$ & -\\
            $q_{rs}$ & Travel demand from $r$ to $s$ & $veh/h$\\
            $\rho$  & Penalty & - \\
            $\hbar$ & Planck's constant & $m^2 kg/s$\\
            $\Gamma$ & Mutated value    & - \\
            $\eta$  & Mutation rate & - \\
            $T_k$  & Tournament size & - \\ \hline
        \end{tabular}
    \end{center}
\end{table}
\setlength{\parindent}{.15in} The upper-level problem involves a public authority determining which roadway links in the network should be restored and what proportion of capacity should be recovered for each, given a limited financial budget, after a disaster. The objective of the public authority is to minimize the total recovery deficiency index and inequity measures (Gini index), or a combination of both. Since these two objectives often conflict, the aim is to find a balance by minimizing a weighted sum of the normalized recovery deficiency index and Gini index. \\
\begin{flalign}
\min \quad & R = \mu \cdot D + (1 - \mu) \cdot E, \label{equ:obj1} \\
\text{s.t.} \quad & \sum_{a\in A} M_a (C_a^1) \le B, \label{equ:const1} \\
& C_a^1 \geq 0, \quad \forall a \in A, \label{equ:const2} \\
& C_a^0 + C_a^1 \leq C_a, \quad \forall a \in A. \label{equ:const3}
\end{flalign}
\setlength{\parindent}{.15in} The variable \textit{E} represents inequity by utilizing the Gini coefficient, which considers income levels. 
\begin{equation}
    E = \frac{1}{2N^2\Bar{I}} \sum_{r\in N} \sum_{s \in N} |I_r - I_s|, \label{equ:equity}
\end{equation} 
where $N$ is the total number of zones in the neighborhood, $I_r$ and $I_s$ are the income levels of neighborhoods $r$ and $s$. $\Bar{I}$ is the average income across the whole network: 
\begin{equation}
    \Bar{I} = \frac{\sum_r I_r}{N},
\end{equation}

To calculate equity in the network, nodes are categorized into low, average, and high-income groups based on the city’s average income. The low and high-income levels are approximated as 60\% and 150\% of the average income, respectively. This results in normalized values of 0.6 for low income, 1.0 for average income, and 1.5 for high income. These normalized values are applied in the equity calculation formula to represent income disparities across the network, enabling a standardized analysis that takes into account the relative economic differences among nodes.
$D$ expresses the recovery deficiency index before and after the restoration decision, defined as:
\begin{equation}
    D = \frac{T_0 - T_1}{ T_0}, \label{d}
\end{equation} 
where $T_0$ and $T_1$ represent the total system travel times before and after the restoration, respectively, and can be computed as follows:
\begin{equation}
    T_0 = \sum_{a \in A} x_a \cdot t_a (x_a, C_a), \label{t0}
\end{equation} 

\begin{equation}
    T_1 = \sum_{a \in A} x_a \cdot t_a(x_a, C_a^0 + C_a^1), \label{t1}
\end{equation} 
where $t_a(x_a, C_a)$ is the travel time on link $a$ with a capacity $C_a$ before the disaster and is defined by the Bureau of Public Roads (BPR) as follows:
\begin{equation}
    t_a(x_a, C_a) = t_0\Bigg(1 + \alpha\bigg(\frac{x_a}{C_a}\bigg)^\beta\Bigg).
\end{equation}
\setlength{\parindent}{.25in} The lower level problem can be formulated as involving road users affected by roadway network capacity degradation or link loss due to the disaster. This impact is quantified using the user equilibrium model for traffic flow assignment:
\begin{flalign}
\min \quad & \sum_a  \int_0^{x_a} t_a(w, C_a^0 + C_a^1) \, dw, \label{equ:obj2} \\
\text{s.t.} \quad & t_a(w, C_a^0 + C_a^1) = t_0 \left( 1 + \alpha \left( \frac{w}{C_a^0 + C_a^1} \right)^\beta \right), \quad  \label{const4} \\
& x_a = \sum_r \sum_s \sum_k f_{rs}^k \delta_{rs}^{a,k}, \quad \forall a \in A, \label{const5} \\
& \sum_k f_{rs}^k = q_{rs}, \quad \forall r, s \in N, \label{const6} \\
& f_{rs}^k \geq 0, \quad \forall k \in K_{rs}, \forall r, s \in N. \label{const7}
\end{flalign}
\setlength{\parindent}{.15in} Equation (\ref{equ:obj1}) represents the objective function where $\mu$ serves as a weight in the restoration process. When $\mu = 0$, the objective function focuses entirely on equity, disregarding the recovery deficiency index. Conversely, when $\mu = 1.0$, the objective function emphasizes only the recovery deficiency index, with no consideration of equity. This weighting allows for a flexible balance between equity and recovery deficiency in the restoration process, $D$ is the recovery deficiency and $E$ is the equity measure. 

Constraint \ref{equ:const1} ensures that the cost of the restoration plan does not exceed the total budget, where $M_a(C_a^1)$ denotes the cost for restoring link $a$ to capacity $C_a^1$ and $M_a(0) = 0, \forall a \in A$. Constraints \ref{equ:const2} and \ref{equ:const3} ensure that the capacity recovery for each candidate link is non-negative and does not exceed the initial capacity before the hurricane. Equation \ref{equ:obj2} defines the objective function for the User Equilibrium (UE) model. Equation \ref{const4} provides the formula to calculate the travel time on link $a$. Constraint \ref{const5} calculates the traffic flow on link $a$ as the sum of all path flows that include link $a$. Constraint \ref{const6} ensures that the sum of path flows from node $r$ to node $s$ equals the demand between these nodes. Lastly, Constraint \ref{const7} ensures that the path flow is non-negative.

\subsection{Study Site and Data} 
\begin{figure*}[!t]
\centering
\subfloat[]{\includegraphics[width=0.31\textwidth]{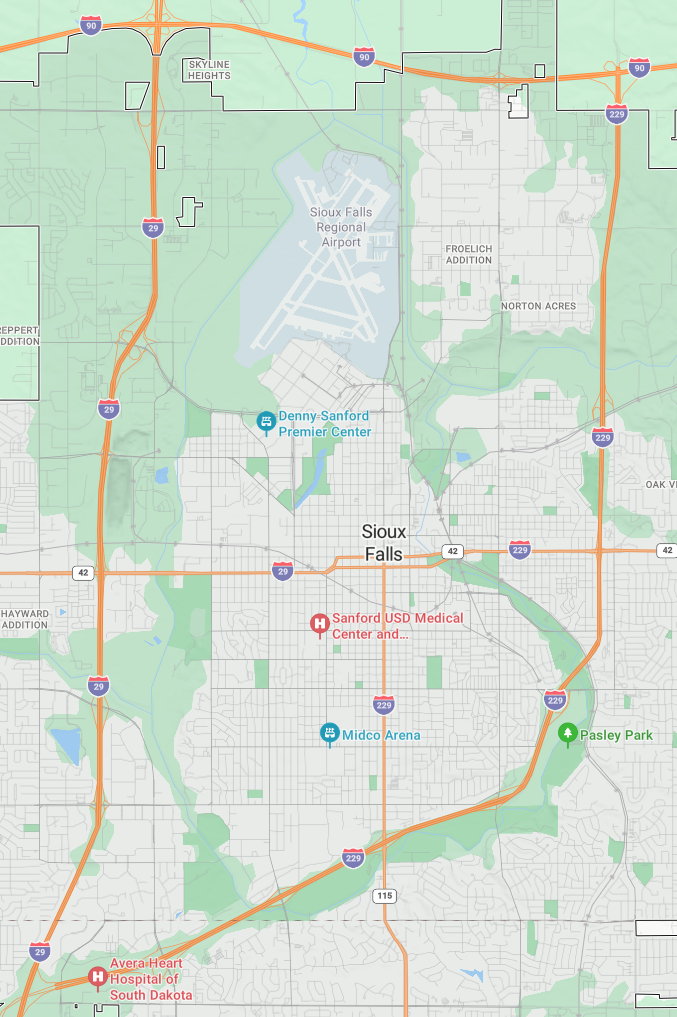}%
\label{fig:net_real}}
\hfil
\subfloat[]{\includegraphics[width=0.4\textwidth]{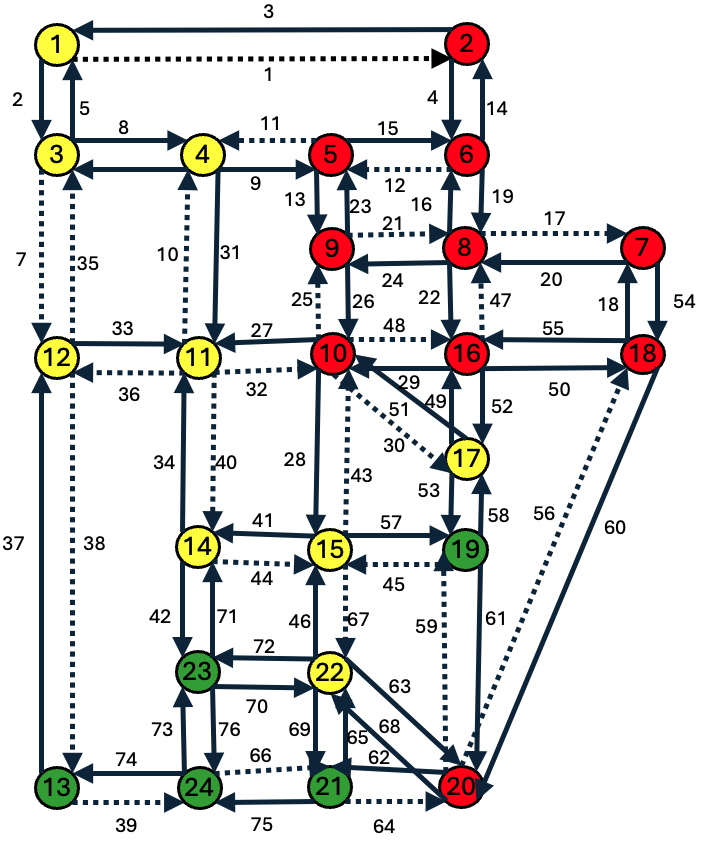}%
\label{fig:net_abs}}
\caption{Sioux Falls transportation network representation (a) Real-world city network (b) Simplified abstract transportation network model}
\label{fig_sim}
\end{figure*}

\setlength\parindent{0pt} Sioux Falls is chosen as a study site due to its unique blend of urban growth, diverse population, and evolving transportation infrastructure \cite{olson2004dakota}. As a mid-sized city with ongoing efforts to expand and modernize its transportation systems, Sioux Falls presents an ideal environment to explore how equitable practices and quantum computing can be integrated into urban planning. The city's initiatives around sustainability, accessibility, and economic development provide a valuable framework for analyzing the restoration and improvement of transportation networks in a way that benefits all community members, particularly underrepresented and marginalized populations. Figure \ref{fig:net_real} illustrates the fundamental transportation infrastructure, depicting a network modeled on Sioux Falls. The network includes 24 zones, each represented by a node in Figure \ref{fig:net_abs}, with each node serving as the focal point for a specific area within the community. The demand across the network is assumed to be fixed as presented in \cite{lu2016optimal}, while the link parameters are given in Table \ref{tab:link_capacity} as reported in \cite{wang2017modified, zhao2020transportation}. 

Additionally, the median per capita income of Sioux Falls, South Dakota, is approximately \$39,293\footnote{https://bestneighborhood.org/per-capita-income-sioux-falls-sd}. Based on this data, an abstracted network map has been created to represent low-, average-, and high-income areas within the city. As shown in Figure \ref{fig:net_abs}, nodes in red signify low-income areas, while nodes in yellow and green represent average- and high-income areas, respectively. This income-based node categorization is utilized to incorporate equity considerations in the restoration of the transportation links. By assigning priority based on the income levels of the areas served by each link, the model aims to achieve a balanced and equitable restoration process across all regions. This income-based approach aligns transportation restoration with social equity goals by considering the needs of economically vulnerable communities in the planning and implementation phases.
\\

\setlength{\parindent}{.15in} For simplicity, the experiment assumes that up to 25 links may be damaged, represented by dotted lines in Figure \ref{fig:net_abs}. The constants used in the travel time calculations in Equation (\ref{const4}) are set as $\alpha = 0.15$, $ \beta = 4$. Additionally, it is assumed that the demand is the same before and after the disaster and that demand cannot exceed supply. It is also assumed that the cost to restore 1 unit of capacity is uniform across all links. As a result, the budget constraint in Equation \ref{equ:const1} is equivalent to the capacity assignment constraint:

\begin{equation}
    \sum_{a\in A}C_a^1 \leq C,
\end{equation}
\\
where \textit{C} is the maximum capacity recovery size, a proxy of monetary budget. The free-flow travel time and capacity for different links are provided and summarized in Table \ref{tab:link_capacity}.

\begin{table}[!t]
    \centering
    \tabcolsep 2pt
    \caption{Original link capacity and free-flow travel time of the links across the network}
        \begin{tabular}{ c c c c c c c } \hline
        Link  & Free-flow   & Capacity  & Link & Free-flow  & Capacity  \\ 
        Number  &  travel time     &  ($10^3 veh/h$)     & Number    &  travel time     & ($10^3 veh/h$) \\ 
        &  (min)     &     &      &   (min)   &\\\hline
        1  & 3.60 & 6.02  & 39 & 2.40 & 10.18 \\ 
        2  & 2.40 & 9.01  & 40 & 2.40 & 9.75  \\ 
        3  & 3.60 & 12.02 & 41 & 3.00 & 10.26 \\ 
        4  & 3.00 & 15.92 & 42 & 2.40 & 9.85  \\ 
        5  & 2.40 & 46.81 & 43 & 3.60 & 27.02 \\ 
        6  & 2.40 & 34.22 & 44 & 3.00 & 10.26 \\
        7  & 2.40 & 46.81 & 45 & 2.40 & 9.64  \\ 
        8  & 2.40 & 25.82 & 46 & 2.40 & 20.63 \\
        9  & 1.20 & 28.25 & 47 & 3.00 & 10.09 \\ 
        10 & 3.60 & 9.04  & 48 & 3.00 & 10.27 \\
        11 & 1.20 & 46.85 & 49 & 1.20 & 10.46 \\ 
        12 & 2.40 & 13.86 & 50 & 1.80 & 39.36 \\ 
        13 & 3.00 & 10.52 & 51 & 4.20 & 9.99  \\ 
        14 & 3.00 & 9.92  & 52 & 1.20 & 10.46 \\ 
        15 & 2.40 & 9.90  & 53 & 1.20 & 9.65  \\ 
        16 & 1.20 & 21.62 & 54 & 1.20 & 46.81 \\ 
        17 & 1.80 & 15.68 & 55 & 1.80 & 39.36 \\ 
        18 & 1.20 & 46.81 & 56 & 2.40 & 8.11  \\
        19 & 1.20 & 9.80  & 57 & 2.40 & 4.42  \\
        20 & 1.80 & 15.68 & 58 & 1.20 & 9.65  \\ 
        21 & 2.00 & 10.10 & 59 & 2.40 & 10.01 \\
        22 & 3.00 & 10.09 & 60 & 2.40 & 8.11  \\
        23 & 3.00 & 20.00 & 61 & 2.40 & 6.05  \\ 
        24 & 2.00 & 10.10 & 62 & 3.60 & 10.12 \\ 
        25 & 1.80 & 27.83 & 63 & 3.00 & 10.15 \\ 
        26 & 1.80 & 27.83 & 64 & 3.60 & 10.12 \\ 
        27 & 3.00 & 20.00 & 65 & 1.20 & 10.46 \\ 
        28 & 3.60 & 27.02 & 66 & 1.80 & 9.77  \\ 
        29 & 3.00 & 10.27 & 67 & 2.40 & 20.63 \\ 
        30 & 4.20 & 9.99  & 68 & 3.00 & 10.15 \\ 
        31 & 3.60 & 9.82  & 69 & 1.20 & 10.46 \\ 
        32 & 3.00 & 20.00 & 70 & 2.40 & 10.00 \\ 
        33 & 3.60 & 9.82  & 71 & 2.40 & 9.85  \\ 
        34 & 2.40 & 9.75  & 72 & 2.40 & 10.00 \\ 
        35 & 2.40 & 46.81 & 73 & 1.20 & 10.16 \\ 
        36 & 3.60 & 9.82  & 74 & 2.40 & 11.38 \\ 
        37 & 1.80 & 51.80 & 75 & 1.80 & 9.77  \\ 
        38 & 1.80 & 51.80 & 76 & 1.20 & 10.16 \\ \hline
        \end{tabular}
    \label{tab:link_capacity}
\end{table}

\subsection{D-Wave's Hybrid Quantum Approach}
\setlength\parindent{0pt}D-Wave provides quantum computing resources through its Leap cloud platform, supported by the Ocean SDK and specialized toolkits for various applications, including scientific research and commercial optimization \cite{johnson2010scalable}. This study utilized D-Wave's Ocean SDK to explore CQMs and evaluate their performance using hybrid solvers, which integrate quantum and classical computation to solve complex optimization problems efficiently.

\setlength{\parindent}{.15in}The Quantum Processing Unit (QPU), featuring a Pegasus graph topology, improves connectivity and streamlines the embedding process, enabling greater flexibility in mapping problem variables onto the QPU. Tools like the Ocean SDK automate minor-embedding, enabling users to focus on problem formulation rather than hardware constraints \cite{boothby2020next}. Hybrid solvers further improve this process by partitioning CQMs between quantum and classical resources, managing constraints, and scaling to larger solution spaces. These advancements, combined with iterative refinement techniques, leverage the unique strengths of quantum annealing to efficiently explore the solution space for modern optimization challenges.

\textbf{Formulation of the problem as a CQM\\}
To formulate the original problem represented in equations (\ref{equ:obj1}) - (\ref{equ:const3}) as a CQM for D-Wave’s hybrid solver, equations (\ref{t0}) and (\ref{t1}) are substituted into (\ref{d}), with the assumptions $\beta = 4$ and $\alpha = 0.15$, resulting in equation (\ref{equ:D}).
\begin{align}
D = &\ \frac{\sum_{a \in A} \Bigg(x_a \cdot t_0 + \alpha \cdot x_a \cdot t_0 \cdot \bigg(\frac{x_a}{c_a}\bigg)^\beta \Bigg)}
{\sum_{a \in A} \Bigg(x_a \cdot t_0 + \alpha \cdot x_a \cdot t_0 \cdot \bigg(\frac{x_a}{c_a}\bigg)^\beta \Bigg)} \notag \\
&- \frac{\sum_{a \in A} \Bigg(x_a \cdot t_0 + \alpha \cdot x_a \cdot t_0 \cdot \bigg(\frac{x_a}{c_a^0 + c_a^1}\bigg)^\beta\Bigg)}
{\sum_{a \in A} \Bigg(x_a \cdot t_0 + \alpha \cdot x_a \cdot t_0 \cdot \bigg(\frac{x_a}{c_a}\bigg)^\beta \Bigg)}, \label{equ:D}
\end{align}
\setlength{\parindent}{.15in} This simplifies to:
\begin{equation}
    D = \frac{\sum_{a \in A} \Bigg[\alpha \cdot x_a \cdot t_0 \bigg(\bigg(\frac{x_a}{c_a}\bigg)^\beta - \bigg(\frac{x_a}{c_a^0 + c_a^1}\bigg)^\beta\bigg)\Bigg]}{\sum_{a \in A} \Bigg[x_a \cdot t_0 \bigg(1 + \alpha \cdot \big(\frac{x_a}{c_a}\big)^\beta \bigg) \Bigg]},
\end{equation}

Additionally, the absolute value operation in equation (\ref{equ:equity}) can be replaced with a quadratic equivalent as follows: 

\begin{equation}
    E = \frac{1}{2N^2 \bar{W}\bar{I}} \sum_{r\in N} \sum_{s\in N} u_{r,s},
\end{equation}
where
\begin{equation}
u_{r,s} = (I_r - I_s)^2,
\end{equation}
\setlength{\parindent}{.15in} This representation emphasizes differences quadratically rather than linearly and this modification changes the nature of the model to penalize larger deviations more strongly due to the squaring. 
This implies that the overall objective function given in equation (\ref{equ:obj1}) can be rewritten as follows: 
\begin{align}
R = &\ \mu \cdot \frac{\sum_{a \in A} \Bigg[\alpha \cdot x_a \cdot t_0 \bigg(\bigg(\frac{x_a}{C_a}\bigg)^\beta - \bigg(\frac{x_a}{C_a^0 + C_a^1}\bigg)^\beta\bigg)\Bigg]}{\sum_{a \in A} \Bigg[x_a \cdot t_0 \bigg(1 + \alpha \cdot \big(\frac{x_a}{C_a}\big)^\beta \bigg) \Bigg]} \notag \\
& + \frac{1 - \mu}{2N^2 \bar{W} \bar{I}} \sum_{r,s \in N} u_{r,s}.
\end{align}
\setlength{\parindent}{.15in} This CQM can be expressed as a Hamiltonian $(H(x))$ for D-Wave’s hybrid quantum solver, where quantum annealing is used to find solutions by minimizing the “energy” of the system:
\begin{equation}
    H(x) = H_\text{obj} + H_\text{Q},
\end{equation}
where $H_{obj}$ and $H_Q$ represent the Hamiltonians of the objective function and constraints respectively:
\begin{align}
H_\text{obj} = &\ \mu \cdot \frac{\sum_{a \in A} \Bigg[\alpha \cdot x_a \cdot t_0 \bigg(\bigg(\frac{x_a}{C_a}\bigg)^\beta - \bigg(\frac{x_a}{C_a^0 + C_a^1}\bigg)^\beta\bigg)\Bigg]}{\sum_{a \in A} \Bigg[x_a \cdot t_0 \bigg(1 + \alpha \cdot \big(\frac{x_a}{C_a}\big)^\beta \bigg) \Bigg]} \notag \\
& + \frac{1 - \mu}{2N^2 \bar{W} \bar{I}} \sum_{r,s \in N} u_{r,s}.
\end{align}
\begin{equation}
    H_Q = H_B + H_C,
\end{equation}
$H_B$ and $H_C$ represent the budget and capacity constraints expressed in equations (\ref{equ:const1}) - (\ref{equ:const3})
\begin{equation}
    H_\text{B} = \lambda_1 \cdot \left( \sum_{a \in A} M_a(C_a^1) - B \right)^2,
\end{equation}
\begin{equation}
    H_\text{C} = \lambda_2 \cdot \sum_{a \in A} \max(0, C_a^0 + C_a^1 - C_a)^2,
\end{equation}
where $\lambda_1$ and $\lambda_2$ are penalty weights to ensure the constraints are satisfied. Therefore, the constraint Hamiltonian $(H(Q))$ is given as:
\begin{equation}
\begin{aligned}
H_{Q} = &\ \lambda_1 \cdot \left( \sum_{a \in A} M_a(C_a^1) - B \right)^2 \\
&+ \lambda_2 \cdot \sum_{a \in A} \max(0, C_a^0 + C_a^1 - C_a)^2,
\end{aligned}
\end{equation}

This implies that the overall Hamiltonian is given as:
\begin{align}
H(x) = &\ \mu \cdot \frac{\sum_{a \in A} \Bigg[\alpha \cdot x_a \cdot t_0 \bigg(\bigg(\frac{x_a}{C_a}\bigg)^\beta - \bigg(\frac{x_a}{C_a^0 + C_a^1}\bigg)^\beta\bigg)\Bigg]}{\sum_{a \in A} \Bigg[x_a \cdot t_0 \bigg(1 + \alpha \cdot \big(\frac{x_a}{C_a}\big)^\beta \bigg) \Bigg]} \notag \\
&+ \frac{1 - \mu}{2N^2 \bar{W} \bar{I}} \cdot \sum_{r,s \in N} u_{r,s} \notag \\
&+ \lambda_1 \cdot \left(\sum_{a \in A} M_a(C_a^1) - B \right)^2 \notag \\
&+ \lambda_2 \cdot \sum_{a \in A} \left[\max(0, C_a^0 + C_a^1 - C_a)\right]^2. \label{equ:prob}
\end{align}

The quantum annealing process is fundamentally based on the adiabatic theorem, which asserts that if the Hamiltonian evolves slowly enough and the system starts in the ground state of the initial Hamiltonian $H_{initial}$, it will remain in the ground state of the instantaneous Hamiltonian $H(t)$ throughout the evolution \cite{nikmehr2024quantum}. This principle ensures that the system's quantum state tracks the lowest energy state of the Hamiltonian as it transitions from $H_{initial}$ to the problem Hamiltonian $H_{problem}$, ultimately yielding the optimal solution encoded in the ground state of $H_{problem}$.

The total Hamiltonian during the annealing process is expressed as:

\begin{equation}
    H(t) = A(t)H_{initial} + B(t)H_{problem},
\end{equation}

\setlength\parindent{0pt}where $A(t)$ and $B(t)$ are time-dependent coefficients that control the transition between the initial and problem Hamiltonians represented as $H_{initial}$ and $H_{problem}$ respectively. 

\begin{equation}
    H_{initial} = - \sum_i \Delta \sigma_i^x, 
\end{equation}

\setlength{\parindent}{.15in}The term $\sigma_i^x$ denotes the Pauli-X operator acting on the $i-th$ qubit which causes a qubit to flip between its quantum states ($|0\rangle$ and $|1\rangle$), while $\Delta$ specifies the strength of the transverse field applied to the qubits. Consequently, the expression $-\sum_i\Delta \sigma_i^x$ represents a transverse field component in the Hamiltonian.

$H_{problem}$ represents the Hamiltonian encoding the optimization problem, which in this case is given is equation (\ref{equ:prob}).

\begin{align}
H_{problem} = H(x).
\end{align}

The annealing schedule, determined by $A(t)$ and $B(t)$, ensures a gradual transition as follows:
\begin{itemize}
    \item At $t = 0, \ A(0) >> B(0)$, so the system is dominated by $H_{initial}$, ensuring a uniform superposition of all possible states. 
    \item At $t = T, \ A(T) << B(T)$, so the system is dominated by $H_{problem}$, ensuring the ground state corresponds to the optimal solution.
\end{itemize}

\subsection{Genetic Algorithm}
\setlength\parindent{0pt}A genetic algorithm solves problems through selection, crossover, and mutation, evolving a population over generations \cite{huang2016genetic, bettemir2015hybrid, akopov2021improvement}. Selection is performed using a fitness-based approach where the fitness of each individual is calculated using a defined fitness function shown in equation (\ref{ga:fitness}). 
\begin{equation}
   R_j = \mu D_j + (1 - \mu)E_j + \rho_j,  \label{ga:fitness}
\end{equation}
where $D_j$ is the recovery deficiency index for each individual $j$, $E_j$ is the Gini coefficient for individual $j$, and $\rho_j$ is a penalty added if individual $j's$ restoration cost exceeds the budget according to equation (\ref{ga:penalty})
\begin{equation}
    \rho = \begin{cases} 
      (c_a - B) \times \rho, & if \ C_a > B  \label{ga:penalty}, \\
      0, & otherwise,
   \end{cases}
\end{equation}

\setlength{\parindent}{.15in}Selection is carried out with tournament selection in accordance with equation (\ref{ga:selection}) where the selection weights are inversely proportional to the fitness values, meaning individuals with lower fitness (better solutions) have higher selection weights. Two parents are then chosen through weighted random sampling, ensuring that better solutions are more likely to be selected. Additionally, the best individual can be retained to guarantee it moves to the next generation.
\begin{equation}
    \Psi = \operatorname*{argmin}_{j\in T_k}\{R_j\}, \label{ga:selection}
\end{equation}
where $\Psi$ is the selected value, ${T_k}$ is the tournament subset of size $k$ and $R_j$ represents the fitness score of individual $j$.

Crossover is implemented by flattening the parent matrices into one-dimensional arrays and selecting a random crossover point as represented in equation (\ref{ga:crossover1}) and (\ref{ga:crossover2}). The genes before this point are swapped between the two parents to produce two new offspring, which are then reshaped back into matrix form. 
\begin{equation}
    c_1 = [p_1[:p], p_2[p:]], \label{ga:crossover1}
\end{equation}
\begin{equation}
    c_2 = [p_2[:p], p_1[p:]], \label{ga:crossover2}
\end{equation}
where \( c_1 \) and \( c_2 \) represent the offspring (children), \( p_1 \) and \( p_2 \) are the parent solutions, and \( p \) denotes the crossover point where the split and combination of elements occur.

Mutation introduces random variations by altering the restoration sizes within an individual. With a certain probability (mutation rate), a random mutation size is generated, and two indices are selected from the positive restoration sizes. The mutation size is subtracted from one index and added to another, maintaining the overall budget constraint. This process just as illustrated in equation (\ref{ga:mutation}) ensures diversity in the population, preventing premature convergence to local optima.

\begin{equation}
\Gamma_i = 
\begin{cases} 
    r \in [0, r_i], & \text{if } p < \eta, \label{ga:mutation}\\
    x_i, & \text{otherwise}, 
\end{cases}
\end{equation}

In this context, \(r\) represents a random value in the range \([0, r_i]\), where \(r_i\) is the restoration value of the \(i\)-th individual. The variable \(p\) denotes a random probability, \(\eta\) is the mutation rate, and \(x_i\) corresponds to the \(i\)-th individual in the population.

Equation (\ref{ga:budget}) enforces the budget during the optimization, ensuring that the mutation does not result in solutions that violate the problem's resource limits.
\begin{equation}
    \Gamma_i \leftarrow \Gamma_i \times \bigg(\frac{B}{C_a}\bigg). \label{ga:budget}
\end{equation}
where $\Gamma$ represents the mutated value. The genetic algorithm was configured with a population size of 50, a mutation rate of 0.1, a tournament size of 3, and a penalty multiplier of 7000.

\section{EXPERIMENTAL RESULTS} \label{results}
\setlength\parindent{0pt}This section presents the results obtained using D-Wave's hybrid quantum computing solvers as well as the Genetic Algorithm. For simplicity, a single value of $\mu = 0.2$ is used in the resilience measure, while simulations are conducted across various budget levels of 75, 150, 225, and 300. The experimental values from these simulations are recorded and analyzed. Furthermore, equity in the restoration process is examined across different values of $\mu$.

\subsection{D-Wave's Hybrid Solver Results}
\setlength\parindent{0pt}The optimum recovery values for each of the damaged links obtained from D-Wave's hybrid QPU solver is shown in \setlength{\parindent}{.15in}Table \ref{tab:recov-cap}.

\begin{table*}[!t]
\centering
\caption{Recovery capacities of road links for $\mu = 0.5$ across various budget levels, evaluated using: (a) the Genetic Algorithm solver and (b) D-Wave's hybrid solver.}
\subfloat[]{        
\begin{tabular}{c c c c c c c} \hline
                    &                       & \multicolumn{4}{ c }{Budget levels}\\ \hline
        Link & Maximum recovery  & 75  & 150 & 225 & 300  \\ 
            &  capacity                 &     &     &      &      \\\hline
        1  &  6.01  & 2.73 & 1.16 & 4.78 & 3.56 \\
        7  &  6.81  & 5.04 & 1.49 & 0.72 & 2.07 \\
        11 & 46.85  & 4.30 & 0.64 & 9.87 & 31.53 \\ 
        12 & 9.90   & 1.59 & 2.18 & 9.68 & 8.58 \\
        17 & 15.68  & 4.65 & 8.56 & 8.46 & 4.52 \\
        21 & 10.10  & 0.61 & 1.55 & 6.53 & 6.68 \\ 
        25 & 27.83  & 3.15 & 5.02 & 6.84 & 12.01 \\ 
        48 & 10.27  & 4.96 & 5.33 & 8.83 & 7.38 \\ 
        30 & 9.99   & 2.30 & 7.14 & 9.53 & 9.89  \\
        10 & 9.82   & 0.98 & 6.99 & 5.25 & 9.69 \\ 
        32 & 20.00   & 4.56 & 6.57 & 6.23 & 10.26  \\
        36 & 9.82   & 3.33 & 5.26 & 8.82 & 7.29 \\ 
        40 & 9.75  & 2.69 & 2.84 & 5.68 & 3.94  \\
        35 & 46.81   & 2.72 & 16.51 & 31.59 & 22.95 \\ 
        38 & 51.80  & 6.50 & 32.26 & 41.06 & 44.85 \\
        39 & 10.18   & 0.72 & 6.87 & 2.94 & 9.55 \\ 
        44 & 10.26  & 2.48 & 6.69 & 9.30 & 9.30 \\ 
        43 & 7.02  & 6.11 & 6.35 & 6.92 & 6.92 \\
        67 & 20.63  & 1.93 & 3.72 & 5.17 & 16.29 \\ 
        47 & 10.09  & 2.13 & 3.05 & 4.24 & 8.14  \\
        45 & 4.42   & 0.64 & 2.81 & 3.91 & 1.93  \\
        59 & 6.05  & 1.21 & 2.23 &  3.10 & 5.59 \\ 
        56 & 8.11   & 2.62 & 2.69 & 3.74 & 3.39 \\ 
        64 & 10.12  & 1.63 & 0.97 & 1.35 & 0.26 \\ 
        66 & 9.77   & 2.78 & 3.04 & 4.23 & 8.96 \\ \hline
      Sum  & 388.09  & 72.38 & 141.92 & 208.77 & 255.51 \\ \hline
    \end{tabular}\label{tab:xxxx}}
\hfil
\subfloat[]
{ 
\begin{tabular}{ c c c c c c c } \hline
                    &                       & \multicolumn{4}{ c }{Budget levels}\\ \hline
        Link & Maximum recovery  & 75  & 150 & 225 & 300  \\ 
             & capacity                 &    &      &    &   \\\hline
        1  &  6.01  & 0.00 & 0.00 & 0.00 & 0.00 \\
        7  &  6.81  & 0.00 & 0.00 & 0.00 & 0.00 \\
        11 & 46.85  & 0.00 & 0.00 & 0.00 & 0.00 \\ 
        12 & 9.90   & 0.00 & 0.00 & 0.00 & 0.00 \\
        17 & 15.68  & 0.00 & 0.00 & 0.00 & 0.00 \\
        21 & 10.10  & 0.00 & 0.00 & 0.00 & 7.26 \\ 
        25 & 27.83  & 0.00 & 0.00 & 0.00 & 27.83 \\ 
        48 & 10.27  & 0.00 & 0.00 & 0.00 & 10.27 \\ 
        30 & 9.99   & 0.00 & 0.00 & 0.00 & 9.99  \\
        10 & 9.82   & 0.00 & 0.00 & 0.00 & 9.82 \\ 
        32 & 20.00   & 0.00 & 0.00 & 10.17 & 20.00  \\
        36 & 9.82   & 0.00 & 0.00 & 9.82 & 9.82 \\ 
        40 & 9.75  & 0.00 & 0.00 & 9.75 & 9.75  \\
        35 & 46.81   & 0.00 & 1.55 & 46.81 & 46.81 \\ 
        38 & 51.80  & 0.00 & 51.80 & 51.80 & 51.80 \\
        39 & 10.18   & 0.00 & 10.18 & 10.18 & 10.18 \\ 
        44 & 10.26  & 0.00 & 10.26 & 10.26 & 10.26 \\ 
        43 & 7.02  & 5.81 & 7.02 & 7.02 & 7.02 \\
        67 & 20.63  & 20.63 & 20.63 & 20.63 & 20.63 \\ 
        47 & 10.09  & 10.09 & 10.09 & 10.09 & 10.09  \\
        45 & 4.42   & 4.42 & 4.42 & 4.42 & 4.42  \\
        59 & 6.05  & 6.05 & 6.05 & 6.05 & 6.05 \\ 
        56 & 8.11   & 8.11 & 8.11 & 8.11 & 8.11 \\ 
        64 & 10.12  & 10.12 & 10.12 & 10.12 & 10.12 \\ 
        66 & 9.77   & 9.77 & 0.00 & 0.00 & 9.77 \\  \hline
      Sum  & 388.09  & 75.00 & 140.23 & 215.23 & 300.00 \\ \hline
    \end{tabular}\label{tab:yyyy}}

\label{tab:recov-cap}
\end{table*}

Table \ref{tab:recov-cap}(b) demonstrates the solver's capacity to optimize link recovery under varying budget levels, highlighting its ability to strategically allocate resources to maximize recovery impact. At the lowest budget of 75, the solver prioritizes highly impactful links, such as Links 43, 47, 59, and 67, which together yield a recovery value of 75.00. As the budget increases to 150, additional critical links, such as Links 38 and 39, are restored, resulting in a cumulative recovery value of 140.23. With a budget of 225, broader recovery is achieved, including the full restoration of Links 32, 35, and 40, further boosting the network’s capacity to 215.23. At the highest budget level of 300, the solver maximizes recovery by addressing all previously restored links and allocating resources to Links 21, 25, and 48, achieving a total recovery of 300.00. Notably, the solver demonstrates efficient resource utilization by aligning recovery values closely with the available budget at each level. This optimized allocation ensures that critical segments are prioritized for restoration, significantly enhancing the network's resilience under financial constraints.

\begin{figure*}[!t]
\centering
\subfloat[]{\includegraphics[width=2.5in]{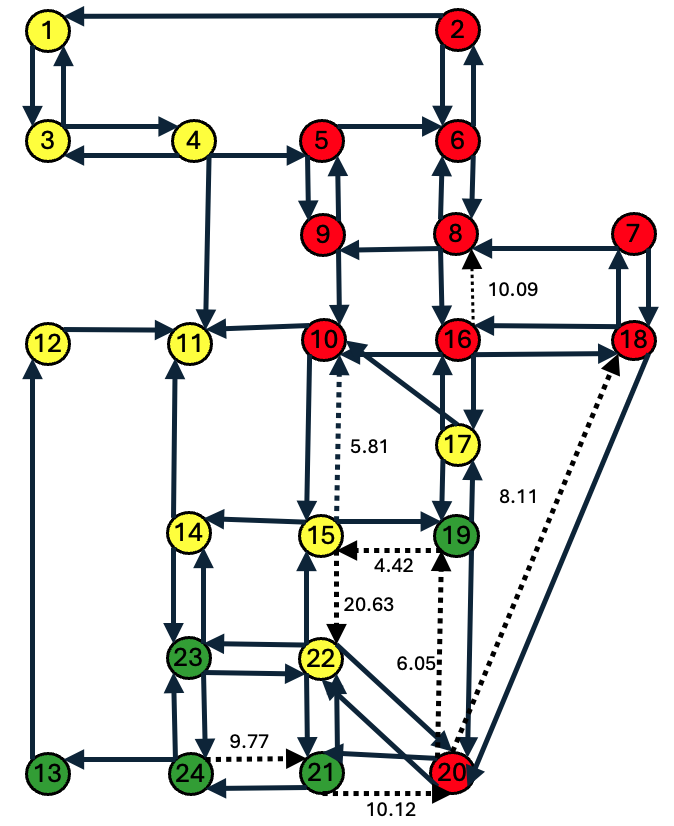}%
\label{fig_first_case}}
\hfil
\subfloat[]{\includegraphics[width=2.5in]{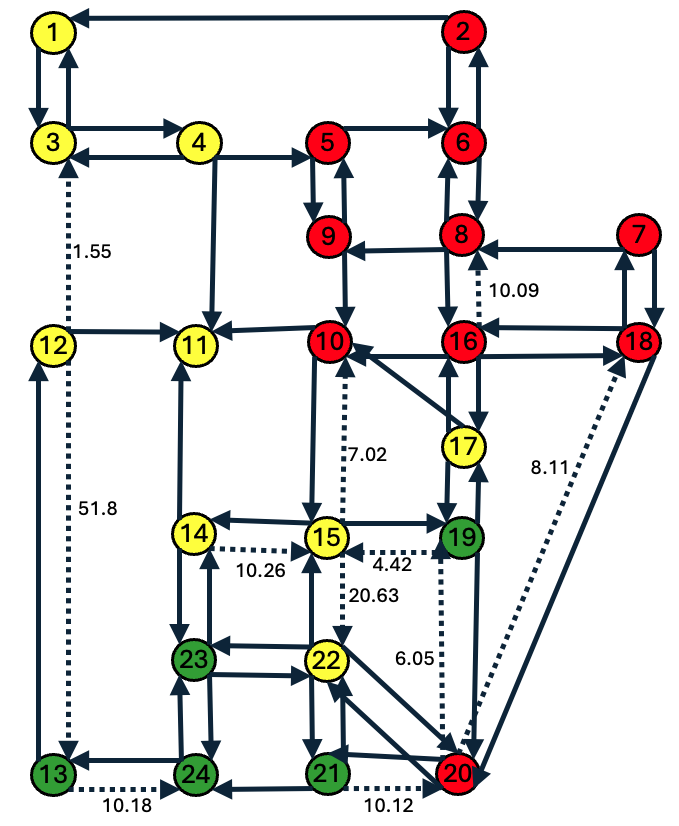}%
\label{fig_second_case}}

\subfloat[]{\includegraphics[width=2.5in]{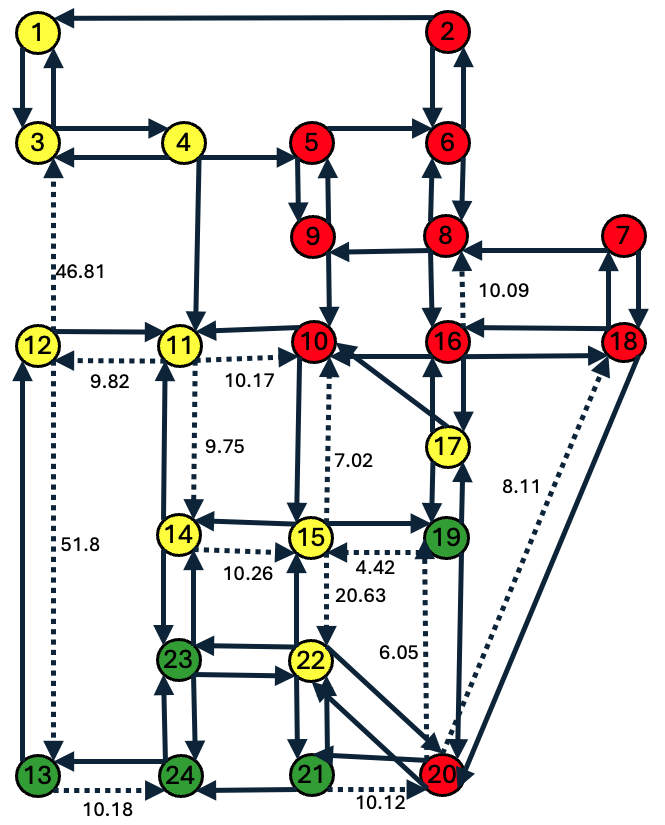}%
\label{fig_first_case}}
\hfil
\subfloat[]{\includegraphics[width=2.5in]{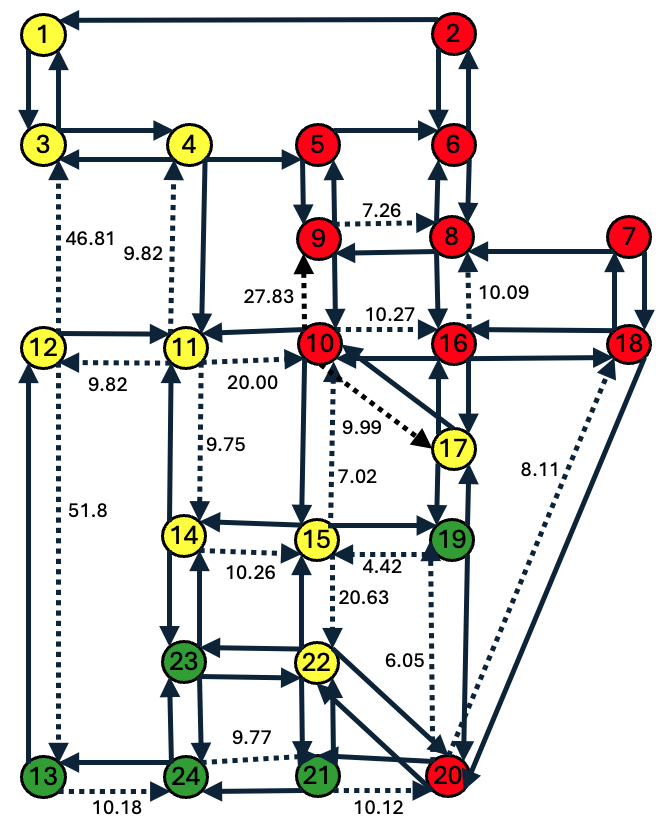}%
\label{fig_second_case}}
\caption{Restored link capacities for different budget sizes: (a) Budget size = 75 (b) Budget size = 150 (c) Budget size = 225 (d) Budget size = 300}
\label{fig:rest_map}
\end{figure*}

In Figure \ref{fig:rest_map}, the progressive restoration of transportation links under different budget constraints is illustrated, with a clear focus on prioritizing connections that serve low-income areas (red nodes) before expanding to average (yellow nodes) and high-income (green nodes) areas. The initial state and low-budget restorations show targeted efforts to maintain essential connectivity for vulnerable populations. As budgets increase, the restoration spreads to include more links, enhancing overall network robustness and accessibility. With the highest budgets, near-complete restoration is achieved, supporting comprehensive network resilience. This incremental approach ensures equitable resource allocation, focusing on restoring critical routes for those most in need and progressively expanding to provide balanced connectivity across all income groups.

\begin{figure*}[!t]
\centering
\subfloat[]{\includegraphics[width=3.5in]{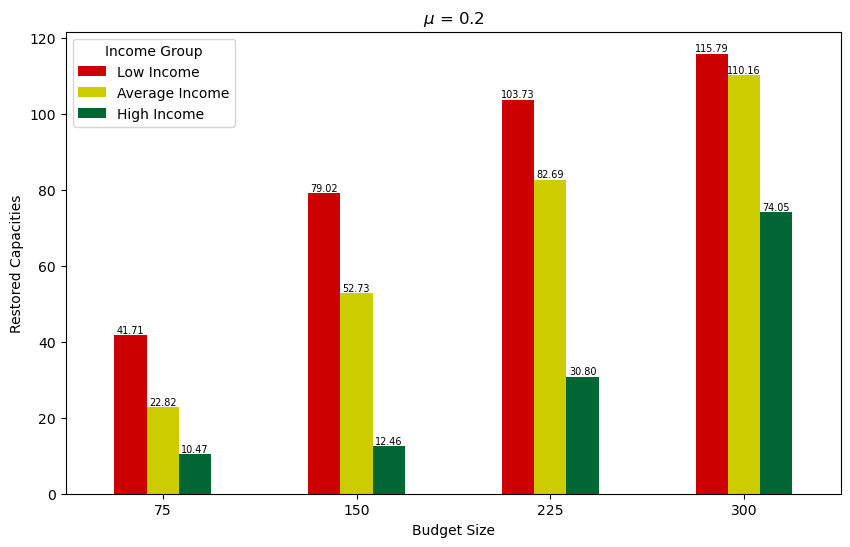}%
\label{fig:equitycoef}}
\hfil
\subfloat[]{\includegraphics[width=3.5in]{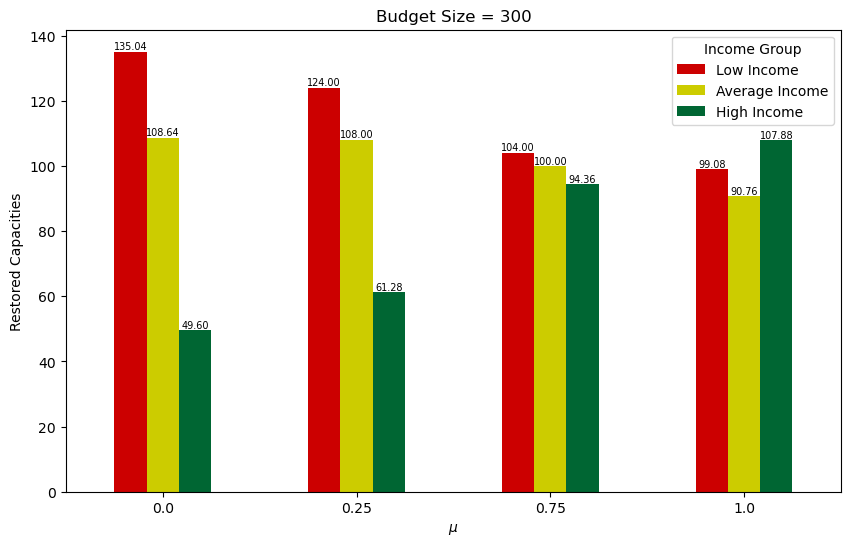}%
\label{fig:equitycoef2}}
\caption{Comparison of restored capacities  across income groups: (a) Under different budget sizes (b) Across varying values of $\mu$}
\label{fig_sim}
\end{figure*}

Figure \ref{fig:equitycoef} highlights disparities in restored capacities for links serving low-income, average-income, and high-income areas across different budget sizes, raising key equity considerations. For all budget levels, links serving low-income areas receive consistently higher restored capacities compared to those in average and high-income areas. This prioritization reflects an equitable restoration approach, ensuring that resources are allocated to communities that may be more vulnerable and reliant on accessible transportation. Average-income areas also benefit from significant restoration, particularly at higher budget levels, though their recovery levels generally remain below those of low-income areas. High-income areas receive the least restoration across all budgets, indicating a de-prioritization based on potentially higher resilience or alternative resources available to these communities. This pattern demonstrates a commitment to equity, addressing transportation needs where they are most critical and enhancing network accessibility for lower-income groups as budget allowances expand.

The percentage changes relative to the 75-budget further illustrate these equity dynamics. For low-income areas, restored capacities increase sharply with each budget increment, rising by 89.45\% at 150, 148.69\% at 225, and peaking at 177.61\% at the 300-budget level. This substantial increase highlights a strong prioritization of low-income areas as budgets grow, likely driven by the higher transit needs and vulnerabilities in these communities. Average-income areas experience even more pronounced increases, particularly at higher budget levels, with restored capacities rising by 131.07\% at 150, 262.36\% at 225, and an impressive 382.73\% at 300. These results suggest that, while initially lower in priority than low-income areas, average-income communities receive increasing support as additional resources become available, reflecting a progressive allocation strategy.

In contrast, high-income areas exhibit more modest but steady growth, with percentage changes of 19.01\% at 150, 194.17\% at 225, and a significant 607.26\% at 300. This restrained restoration growth at lower budget levels underscores the equity-oriented approach, ensuring that restoration efforts focus on areas with greater needs. Overall, these findings demonstrate a deliberate and structured approach to equitable restoration. Increasing budget allocations amplify the benefits for low- and average-income communities, ensuring a fair and impactful distribution of restored capacities while addressing disparities and prioritizing the most vulnerable populations.

Figure \ref{fig:equitycoef2} illustrates the restored transportation capacities for different income groups (low, average, and high) under a budget size of 300, with varying levels of the equity-focused weight factor $\mu$. As $\mu$ increases from 0.0 to 1.0, the restored capacities for low- and average-income areas generally decrease, while the restored capacities for high-income areas show a significant increase. This shift reflects the model's adjustment in prioritization, transitioning from a focus on equity at lower $\mu$ values to a more balanced restoration across income groups as $\mu$ rises.

The percentage changes in restored capacities relative to $\mu = 0.0$ provide further insight into this balance. For low-income areas, restored capacities decrease steadily with higher $\mu$ values, showing a decline of 8.18\% at $\mu = 0.25$, 22.99\% at $\mu = 0.75$, and 26.63\% at $\mu = 1.0$. These reductions suggest a gradual shift away from prioritizing low-income areas as the weight factor shifts toward broader efficiency considerations.

Average-income areas show a less pronounced decline, with a small decrease of 0.59\% at $\mu = 0.25$, 7.95\% at $\mu = 0.75$, and 16.46\% at $\mu = 1.0$. While these reductions are moderate compared to low-income areas, they indicate that average-income communities are still deprioritized to some extent under higher $\mu$ values.

In contrast, high-income areas experience substantial increases in restored capacities as $\mu$ rises. Restored capacities increase by 23.55\% at $\mu = 0.25$, 90.24\% at $\mu = 0.75$, and a remarkable 117.50\% at $\mu = 1.0$. This dramatic growth underscores the model’s flexibility to reallocate resources toward high-income areas as equity considerations diminish.

Overall, the results highlight the model's adaptability. Lower $\mu$ values focus restoration efforts on low- and average-income areas, aligning with equity goals, while higher $\mu$ values prioritize a more even distribution of restored capacities, balancing equity and efficiency.

\begin{table}[!t]
    \caption{Computational times across the hybrid solver experiments}\label{tab:hqpuqTime}
    \begin{center}
    \begin{tabular}{cc} \hline
        Budget Level & Computational time $(s)$ \\ \hline
        75 & 8.749704\\
        150 & 8.750424\\
        225 & 8.748665\\
        300 & 8.748733\\ \hline
    \end{tabular}
    \end{center}
\end{table}

The quantum hybrid solver shows potential advantages over traditional classical methods in terms of computational efficiency and scalability for complex optimization problems like transportation network recovery. While classical methods often experience increased computational time as problem complexity and constraints grow, the quantum approach maintains relatively consistent processing times across various budget sizes (around 8.7 seconds as shown in Table \ref{tab:hqpuqTime}). This demonstrates the quantum method’s ability to handle complex, constraint-heavy scenarios more efficiently, providing faster and potentially more optimal solutions compared to classical techniques, which often require simplifications or suffer from longer processing times in such cases.

\subsection{Genetic Algorithm Results}
\setlength\parindent{0pt}The results obtained by the GA solver, as shown in Table \ref{tab:recov-cap}(a), highlight key differences compared to the hybrid quantum solver, particularly in budget utilization and recovery prioritization across network links. The GA solver demonstrates inconsistent restoration patterns, with partial allocations distributed more evenly across links without a clear focus on high-need areas. At lower budgets, the GA solver tends to allocate small restoration capacities to multiple links, spreading resources thinly instead of concentrating on critical segments, resulting in a more distributed yet lower-impact restoration approach.

\setlength{\parindent}{.15in}At higher budgets, while the GA solver increases recovery capacities, it still does not consistently prioritize links of higher importance or those with higher maximum recovery capacities. This lack of targeted allocation can lead to suboptimal equity outcomes, as the solver does not reliably focus on links serving low-income or vulnerable populations. In contrast, the hybrid quantum solver explicitly prioritizes these areas, producing more socially equitable results. The GA’s restoration totals, such as 72.38 at a budget of 75 and 208.77 at a budget of 225, reveal a somewhat inefficient use of the available budget, particularly at higher levels. This inefficiency and the non-prioritization of equity stem from the GA solver’s inadequate exploration of the search space. However, with algorithm adjustments and tuning, the GA’s performance could be improved, albeit with increased computational time. Ultimately, the GA solver provides a broader, less targeted restoration approach, lacking the equity-focused and efficient optimization achieved by the hybrid quantum solver. The computational times obtained during the experiments across various budget levels is shown in Table \ref{tab:gaTime}

\begin{table}[!t]
    \caption{Computational times across the genetic algorithm experiments}\label{tab:gaTime}
    \begin{center}
    \begin{tabular}{cc} \hline
        Budget level & Computational time $(s)$ \\ \hline
        75 & 664.9874\\
        150 & 665.8943 \\
        225 & 677.6981\\
        300 & 674.6735\\ \hline
    \end{tabular}
    \end{center}
\end{table}

\subsection{Comparison of Computational Mechanisms: Hybrid Quantum Solver vs. Genetic Algorithm}
\setlength\parindent{0pt}The experiments were conducted on a MacBook Pro (2024 model) equipped with an Apple M4 Max chip featuring a 14-core CPU and a 32-core GPU, 36 GB of unified memory, and a 1 TB SSD for storage. The efficiency of the hybrid quantum solver is evident in its consistent computational times across all experiments, as shown in Table \ref{tab:hqpuqTime}. The solver achieved runtimes of approximately 8.7 seconds for budget sizes of 75, 150, 225, and 300, with only marginal variations. These consistent runtimes highlight the hybrid solver’s ability to rapidly solve complex optimization problems irrespective of budget size, reflecting its robustness and scalability.

\setlength{\parindent}{.15in}In contrast, the GA solver relies on evolutionary principles, including selection, crossover, and mutation, to iteratively evolve a population of candidate solutions toward optimality. While this approach is versatile and robust for a wide range of problems, it can be computationally intensive, as it requires evaluating the fitness of multiple solutions over numerous generations. The GA's performance is often bottlenecked by the combinatorial explosion of possibilities in large solution spaces, particularly when constraints and dependencies are complex.

The efficiency of the hybrid quantum solver stems from its ability to process a significant portion of the computational workload using quantum annealing, which is inherently parallel and excels at rapidly finding low-energy configurations in rugged optimization landscapes. Quantum tunneling allows the solver to bypass high-energy barriers that would trap classical solvers in suboptimal solutions, and the classical components ensure that constraints are satisfied and solutions are practical. Additionally, the hybrid approach minimizes the overhead associated with exhaustive search or population-based exploration, making it particularly effective for problems with intricate constraints and large-scale solution spaces.

\section{CONCLUSION} \label{conclusion}
\setlength\parindent{0pt}This study evaluates the effectiveness of hybrid quantum and GA solvers for equitable restoration of transportation networks, particularly in post-disaster scenarios. The hybrid quantum solver proved to be superior in both prioritizing resources for vulnerable, low-income areas and maximizing resilience across various budget levels. With each budget increase, the quantum solver strategically enhanced connectivity for low-income areas while gradually extending benefits to average- and high-income areas. This equity-focused allocation approach ensured that critical links for disadvantaged communities were prioritized, offering a more balanced and socially conscious restoration process. In contrast, the GA solver showed a more diffuse restoration pattern, allocating resources inconsistently across income groups. While it provided some restoration for all areas, the GA approach lacked the targeted focus needed to maximize support for low-income areas, potentially limiting network resilience.

\setlength{\parindent}{.15in}Additionally, the quantum solver demonstrated a significant computational advantage, consistently completing optimization tasks within approximately 8.7 seconds, regardless of budget size, compared to over 600 seconds required by the GA solver. This efficiency makes the hybrid quantum approach especially suitable for urgent recovery scenarios, where swift, effective decision-making is critical. The findings indicate that hybrid quantum computing not only enables rapid, scalable solutions but also aligns with equity goals by focusing on the restoration needs of the most vulnerable communities. As quantum technology continues to advance, its application in network optimization holds great promise for improving urban resilience and addressing complex restoration challenges, ultimately supporting more equitable infrastructure planning.

\bibliographystyle{IEEEtran}
\bibliography{bibliography}

\end{document}